\documentclass{interact}
\pdfoutput=1
\usepackage[utf8]{inputenc}
\usepackage[english]{babel}
\usepackage{hyperref} 
\usepackage{amsmath}
\usepackage{csquotes}
\usepackage{xurl}
\usepackage{lmodern}
\usepackage[hang]{footmisc} 
\usepackage[style=apa]{biblatex}
\begin{document}

\title{The Method of Critical AI Studies, A Propaedeutic}

\author{
\name{Fabian Offert\textsuperscript{a*} and Ranjodh Singh Dhaliwal\textsuperscript{b*}}
\affil{\textsuperscript{a}University of California, Santa Barbara; \textsuperscript{b}University of Basel
\newline
*Equal contributions } }

\maketitle

\renewcommand\abstractname{ABSTRACT}
\begin{abstract}
We outline some common methodological issues in the field of critical AI studies, including a tendency to overestimate the explanatory power of individual samples (the benchmark casuistry), a dependency on theoretical frameworks derived from earlier conceptualizations of computation (the black box casuistry), and a preoccupation with a cause-and-effect model of algorithmic harm (the stack casuistry). In the face of these issues,  we call for, and point towards, a future set of methodologies that might take into account existing strengths in the humanistic close analysis of cultural objects.
\end{abstract}

\section*{Introduction}

We begin with a simple question: When is Generative AI `wrong'? If you type in a prompt and the output on your screen makes no sense or is factually incorrect, does that make ChatGPT `wrong'? Perhaps it does. But what if you input the exact same prompt and are confronted with a different answer (as is often the case with probabilistic systems such as contemporary large language models). Was your initial assessment (that ChatGPT is wrong) wrong? Or is the GPT-4 model which forms the backend of ChatGPT wrong? And that is to ask nothing yet of the system's training data, internal representations, and  pre-training, of the system prompts, its interface, or of the system's implementation details.

We should note that the aforementioned scenario does not ask if Generative AI is `bad,' `harmful,' or `bullshit,' all questions asked (and answered, often in the affirmative) by many \autocite{Hogan_2024, Yu_2023, Fischer_2023, Hicks_2024}. Critical AI studies has been ably outlining the various critical positions that are, and need to be, taken by data studies, media studies, technology studies, philosophy, history, gender studies, or studies of culture, aesthetics, politics, and race \autocite{Raley_2023}. What eludes all these (aspirationally) shared projects, however, is a shared methodology. One might even go further and say that what is missing here is any methodological specificity at all: critical AI studies has so far, after all, ported previous methodological gambits from the respective disciplinary configurations and applied it onto the new object/assemblage/practice that is generative AI \autocite{Offert_2025}. This commentary takes, as its starting point, the simple observation that not only is Critical AI, as a project, more vital than ever, but so is the need for a (robust) methodology therein.

Consider the thought exercise with which this commentary began; it is quite likely that one might find such an encounter (with the AI) lying at the heart of some of your favorite AI takedowns. Few things are as satisfying as an supposedly inanimate chimera losing its deceptive powers when confronted with the rational critic; here lies the vanquished demon, slain, for it spewed lies (or worse yet, gibberish) \autocite{Latour_2004}. Specific prompts that generate specific texts, images, or sounds, are often asked, in scholarship and in public discourse today, to stand in for a universal critique of the abilities and possibilities offered by generative AI, a `method' which actually threatens to become increasingly less successful as models continuously improve. Previously one could point, quite literally, at the fingers image generators would add or remove from human hands.\footnote{The necessity of this kind of ``wild forensis" was first established by artist Kyle McDonald in a 2018 blog post \autocites{McDonald_2018}[see also][]{Meyer_2024}.} ``A photograph of a perfect human hand with five fingers" would produce additional fingers because of the model’s inability to parse the prompt as a sentence proper, where ``five fingers" is an intentionally redundant specification of ``perfect human hand", meant to preemptively mitigate the very mistake it then ends up producing. Due to its prominence, this glitch has featured in many recent critiques of image generators, including -- a somewhat sinister twist -- in an academic project on the authenticity of war photography, the one subject matter where the glitch would not necessarily give evidence to some form of image generator working in the background \autocite{Geise_2024}. What is more, the glitch has now been thoroughly fixed in the most recent model generations, all but invalidating the critiques it informs.

Perhaps the easiest way to understand this phenomenon -- an insistence on the evidence status of individual samples -- is by understanding it as a misunderstanding of the idea of `red teaming'. Red teaming is the name given to a practice, in the fields of security and cybersecurity, wherein a group takes on (by prior planning and often consent) the role of an enemy and tries to infiltrate, attack, or harm in other ways, the entity/organization that finally benefits from knowing how its defenses could be breached so as to endeavor improving them. Red teaming is now also frequently used in artificial intelligence by companies who employ hackers and other experts who can point out the problems and vulnerabilities that can be solved, or at least planned for, before the wider world finds out about it. And in fact we are all, as users, implicated in some kind of global red teaming of under-tested AI systems being introduced at a rapid pace.\footnote{It is also worth noting that a similar logic exists within the design of certain AI models, see \textcites{Lepage-Richer_2021, Offert_2021a, Galison_1994}.}

What, then, is the widespread practice of `ask the AI model a question that produces a wrong/unexpected answer' if not `humanities red teaming.' In the spirit of self-critique, we reckon that this development is methodologically specious (or at least weak) and that we can, and must, do better. A somewhat polemical critique, for sure, but one, we argue, that we must subject ourselves to if we want the endless game of catch-up with the technical status quo to come to an end. To be perfectly clear: our argument does not suggest a less critical methodology but the inverse. We ask ``When is Generative AI `wrong'?" not to show that it isn’t but that its failures go much deeper than a handful of examples might suggest.

\section*{How did we end up here?}

How did we end up here? Some of the reasons are entirely clear, and thus not very enlightening. It is  simply fun to hear people talk about their conversation with ChatGPT as if it were something funny that happened to them the other day. Humanists like stories -- to put it mildly -- and everything produced by today’s AI systems has an anecdotal character. The typical customer-servicesque ``certainly, here’s five reasons why…" of ChatGPT, which itself mirrors the buzzfeedification of the Web at large, perfectly lends itself to being sarcastically recounted, especially when the subject matter is serious. Add into the mix the amateurish refusal mandated by corporate censorship and the introductory chapter practically writes itself.

The more interesting reasons, we would like to argue on the following few pages, are historical in their origins and limitations. While we cannot trace in detail the development of a whole field\footnote{See \textcite{Raley_2023} for this kind of overview. See also \textcite{Raley_2024}.} -- which would have to include the development of those fields it heavily intersects with, like science and technology studies, adding to the intractability of the task -- we would like to offer three historical dependencies\footnote{We should remark that our use of language from computer science discourse here, and elsewhere in this piece, is intentional and done following the traditions of science and technology studies where the description and prescription gain added precision by a critical engagement with the material-semiotic configuration being studied.} structuring, and eventually facilitating the methodological omnipresence of the practice of humanities red teaming. Here, we introduce these three as examples of \textit{casuistry}: the rationalization by which a thinker often resolves their own moral dilemmas through the use of inappropriate generalizations and/or applications of theories to specific instances. We are suggesting this term not to belittle the moral challenges that emerge, ever more rapidly, from the real-world use of machine learning systems\footnote{Apart from the well-researched problem of generally biased decision-making, these include, more specifically, high-stakes applications like credit scoring or recidivism prediction, but also `backend' issues like labor violations and environmental harms.} but to point out that, ironically, the tendency of machine learning models to not generalize well has befallen the very scholarship meant to critique this generalization failure.

\section*{The benchmark casuistry}

The first casuistry is the oversized importance ascribed to the Turing Test \autocite{Turing_1950}. It is no secret that the chat interface that dominates today’s models takes its very inspiration from Turing’s original `benchmark' proposal.\footnote{With some historical `stops' in between, for instance Weizenbaum’s \textit{ELIZA}.} As computer scientist Scott Aaronsen maintains: ``one can divide everything that’s been said about artificial intelligence into two categories: the 70\% that’s somewhere in Turing’s paper from 1950, and the 30\% that’s emerged from a half-century of research since then" \autocite{Aaronsen_2013}. While this seems intuitively true\footnote{For instance, Turing’s ``Contrary Views on the Main Question" \autocite[442ff]{Turing_1950} and ``Lady Lovelace’s Objection" \autocite[450f]{Turing_1950} in particular are resurrected every time a machine does  something remotely `creative'. A similar philosophical template exists for the critique of Turing’s position and its derivatives, in the form of Searle’s Chinese Room \autocite{Searle_1980}. Searle’s thought experiment, which famously argues for the context-dependency of meaning (by means of the phenomenological concept of intentionality) is leveraged, time and again, when language models seem to produce `meaning' or exhibit `understanding' – most recently, and perhaps  most prominently, in \textcite{Bender_2020}, where the Chinese Room becomes an octopus on a telephone.}, Turing’s popularity has somewhat obscured the finer details of his argument. And it is exactly these details that already point us to one of our core claims: that probabilistic systems necessarily require a probabilistic form of critique.

``I believe that in about fifty years' time", Turing writes, ``it will be possible to programme computers, with a storage capacity of about $10^9$, to make them play the imitation game so well that an average interrogator will not have more than 70 per cent. chance of making the right identification after five minutes of questioning" \autocite[442]{Turing_1950}. In other words: around the year 2000 a computer with a storage capacity of roughly 120MB should `win' the imitation game with a likelihood of 70:30 after five minutes of playtime. These are quite concrete requirements, and nowhere does Turing indicate that any universal predictions can be generated from this particular one, whereas we can easily verify this particular one (to the degree that our imitation game program works well enough) by averaging a number of five-minute runs across a slate of test subjects. While all of this seems obvious from the quoted passage, this precise implication is nevertheless often left out when Turing is summoned as the common ancestor of artificial intelligence and its critique.

To the computer scientist, this historical contextualization of the Turing Test will be entirely unsurprising. How else, they might say, would one conceive of a benchmark of a system of unknown capabilities if not by means of  probabilistic sampling under fixed external conditions? And indeed, actual AI benchmarks do exactly that, including the thematically (to us) appropriate BIG Bench (Beyond the Imitation Game) framework which provides over 200 tasks that all are measured by running a task several times and noting the likelihood of outcomes \autocite{Ghazal_2013}. And yet, humanities critiques are dominated by $n=1$ experiments, where a single input-output example serves to ‘discredit’ the entire probabilistic system. We argue that this is not exclusively the result of a lack of technical skills or resources, but due to a misaligned application of roughly hermeneutically-inspired methods to probabilistic systems. In other words, humanities scholars may ironically be finding themselves red-handed at the sight of a crime they often accuse the AI of committing: misrecognizing everything as a nail by self-identifying as a hammer.

A few clarifications are in order here. First, this is not a call for considering the general replicability of humanities scholarship as some digital humanities would stipulate it. In fact, the particular idiosyncrasies of humanities scholarship might, in our opinion, give a competitive edge to a properly formulated methodology, as we outline below. Second, we absolutely do not suggest that qualitative methods have no place in critical AI studies. Many studies in computer science have inadvertently demonstrated that the attempt to operationalize away the core philosophical  problems that AI systems pose indeed sabotages critical efficacy. A good – well, bad – example here is the ever recurring attempt to prove that machines `can make art' by letting users pick a `prefered' image from a lineup of two, all but ignoring the discursive, material, and temporal dimensions of works of art, which are presented only as pixelated JPEGs shown on mid-level computer displays in a neon-lit CS lab. Third, and arguably most important: we do not argue that the humanities need to copy technical benchmarks. Rather the humanities, in the context of critical AI studies, need a better grasp on the interface of qualitative and quantitative critique at which it operates. An interesting starting point to lend visibility to this interface might be Johanna Drucker's proposal \autocite{Drucker_2018} to understand the act of interpretation as the collapse of the probability space of all possible interpretations -- to always keep in mind, in other words, that even the best anecdote is a sample, and necessarily reductive \autocite{Simpson_1995}.

\section*{The black box casuistry}

The second historical dependency at work in humanities red teaming is the understanding of cybernetics as the catch-all historical predecessor of artificial intelligence. Clearly, so much of contemporary artificial intelligence research is determined by cybernetics research, and, perhaps more significantly, the military-industrial funding incentives extensively utilized by cybernetics in the 1950s and 1960s \autocites{Dobson_2023, Pasquinelli_2023, Dhaliwal_2024, Phan_2024}. This is not to say -- at all -- that this history is widely known or does not need to be further investigated. Instead, pointing out the cybernetic connection is meant to emphasize that among the very few explicitly methodological inspirations of critical AI studies, the `black box', taken straight from cybernetics, ranks among the most prominent. ``AI models are black boxes", in 2024, sounds like a truism, and could yet not be further from the truth. Yes, AI models are complex systems, and yes, there is no easy way to infer, purely from the weights and biases of a neural network, what the model does, or what data it was trained on. But AI models rarely consist of just a single neural network, nor do they come into the world as entirely new systems, trained on entirely new data, with entirely new mechanisms. AI models are historical, maybe even `more historical' than many other technical objects. Every new model builds on an entire architectural history, a history of how things are done with the parts that are available. Progress in artificial intelligence research, even if the pace of developments seems to suggest otherwise, is fundamentally incremental. Not only do new models build onto the insights gained from previous generations, they often embed parts of these models, or entire models, within the new model  pipeline.

Where does the humanities' fixation on the black box come from, then? As so often, it comes from a development that \textcite{Vinsell_2021} has called `criti-hype': the counterintuitive alignment of hype and critique, where critics require worrisome developments to take place to critique, while corporations utilize them to present their products as more powerful than they actually are. The boldest example of this might be Sam Altman's Senate testimony, a masterclass on corporate propaganda \autocite{Hogan_2023}. And at least to a degree, it is also the field of explainable artificial intelligence (XAI) that requires to uphold the black box narrative to be able to push their (sometimes spurious and quasi-scientific) methods of probing and dissecting neural networks. ``As AI becomes more advanced, humans are challenged to comprehend and retrace how the algorithm came to a result. The whole calculation process is turned into what is commonly referred to as a ``black box" that is impossible to interpret. These black box models are created directly from the data. And, not even the engineers or data scientists who create the algorithm can understand or explain what exactly is happening inside them or how the AI algorithm arrived at a specific result." This quote is not taken from a piece of AI critique but from IBM’s website on explainable artificial intelligence, and it demonstrates the fine line XAI scholarship has to maintain, between upholding the black-box nature of its object of study (such that the tools the development of which is front and center in XAI scholarship) and arguing for the existence of a ``rich inner world" of AI models, worthy of empirical study akin to the natural world \autocites{Olah_2020, Offert_2021b, Offert_2023}.

The fallout of this black-box casuistry then is, again, mainly methodological, or rather, it exacerbates the absence of methodological innovation. If AI systems are black boxes, then probing their inputs and outputs must be enough to arrive at a coherent model of their functioning. The result are folk theories \autocite{Kempton_1986} of artificial intelligence: theories which seem to work in practice (i.e., that seem to give evidence to a critical perspective) but are entirely disconnected from the technical reality of the objects they purport to describe. Akin to theories of social media app surveillance, they often contain traces of technical truth but fail to ground their critique as a whole. And nothing is more devastating to the critic than being labeled a conspiracy theorist instead of a critical one \autocite{Masco_2023}.

Like in the benchmark casuistry described above, the commonly-used chat- or prompt-based interface contributes to the black-box casuistry. The linguistic interaction that forms the lynchpin of these interfaces not only helps sustain some of the aforementioned narrativization in humanities critiques but also encourages this very blackboxing; the simple and largely empty chat window demonstrates no real depth, instead safely ensconcing the technical setup behind a plain void. Perhaps a cue can be taken here from the field of Human-Computer Interaction (HCI). As the name indicates, HCI has always been in the business of understanding humans as a part of the computational assemblage; it can be best understood, in this homological verve, as the traditionally ``humanities" part of computer science. And when HCI as a field, in its internal critiques \autocites{Suh_2024, Kapania_2024, Ma_2024}, is clear about the significant limitations of natural language use in AI interfaces, it should, at the very least, give humanistic criticisms of chat-based interactions some pause. Interfaces, as Alexander Galloway reminds us, are not `windows' into content, but effects produced by technology \autocite{Galloway_2012}. The chat-box seems to have effected, then, a perplexing condition of interpretive illusion for our disciplinary configuration. 

\section*{The stack casuistry}

The final casuistry we invoke engages with a particular genre of algorithmic critique that first emerged in the early 2010s. It can simply be understood thus: the critic's interactions with the sociotechnical system at hand, especially when it breaks (whether that be understood as a glitch, a bug, or a feature), denote, in such accounts, a structural issue for the given algorithm, and often `algorithms' at large. For an enormously influential example of such a method, see Safiya Umoja Noble's landmark book, \textit{Algorithms of Oppression}, and the rhetorical maneuver that the title indicates. Now it is undeniable -- and structural analyses at the level of political economy from various other methods clearly corroborate this -- that the issues at hand, and certainly the ones pointed out by Noble, are structural and ubiquitous to sociotechnical cultures at large; the question being posed by us here simply has to do with how to arrive at that analysis. 

Algorithmic critique assumes a certain operation at the back-end, where $A$ input combined with $B$ computational decision processes leads to $C$ output. Especially when aligned with the black-box casuistry, it simply means understanding the technical system as a linear stack, one where numerous operational units stand atop each other, and poking one of those, like a Jenga block, would bring down (speaking critically) the whole structure. This methodological formation has been well-theorized and does indeed yield remarkable critical results \autocite{Bratton_2015}. However, two caveats must be pointed out here. Firstly, algorithm critique, as it stands, works only for deterministic systems, where one set of inputs, regardless of the user or the context of interaction, yields the same expected output. The regularity and predictability of deterministic systems, once upturned by probabilistic systems such as most contemporary AI -- and more critically, most non-AI internet interactions as well, largely due to the widespread application of aggressive A/B testing on online interfaces and backends over the last decade or so -- means that making any claims about the algorithm from the recounting of an interaction becomes far more tenuous, for the next time you use it, the product may have changed, the test condition might have changed, or you may be interacting with a different sociotechnical system altogether. Secondly, algorithm critique, insofar as it presumes a stable stack underneath to critique via interactions, suddenly finds itself making claims about outdated technical paradigms. The stack has not been ‘updated’ critically, especially in an age of probabilistic systems that require probabilistic critiques. This unlinking of the structure of the object and the structure of its critique is what is at stake in our comments here.

What might rethinking, if not updating, the stack mean in this context? For starters, it would involve engaging seriously with the remnants of the previous form of deterministic algorithmic configuration, carefully going through the Jenga blocks on the ground and noting what remains standing and what has fallen (and when and how) in this newer era of probabilities and extensive test conditions. One might note, for example, that AI models today look different across time, and that the mode and temporality of their internal changes might differ drastically (or mildly) from algorithms of yore. One would also note upon looking closely the issue that `algorithm', insofar as it referred to a strict start and end point definition in computer science \autocite[1--9]{Knuth_1997}, was perhaps already being misapplied in algorithmic critique; by the early 2010s already, algorithms were not temporally delimited as CS would have it, and the word was being colloquially imposed upon a general sense of predictable procedurality.\footnote{The implications of this very point are numerous but beyond the scope of this commentary.} Rethinking the stack might also mean, more importantly, shifting the window of engagement to concepts more suited for the critical purpose. One conceptual shift here, at the very least as a heuristic, might go via contending with 'pipelines' instead of stacks. A pipeline is many things, but referencing the fluid pipelines that carry water, oil, and gas, a pipeline in computation connotes the fluidity, flexibility, infrastructural transitions, and material extraction \autocite{Dhaliwal_2025}. We suggest contending with pipelining not only because it is a conceptual framework better suited to the material, technical conditions of artificial intelligence operations today—data pipelines feed into, and are mixed during the execution cycles of, instruction pipelines, which may themselves be triggered by model pipelining where several training datasets (and the generated weights through training) are brought to  bear, along with several vector databases, towards a multi-step or multi-agentic AI operation—but also because it would more closely match, while still retaining enough critical capacity, a paradigmatic material-semiotic condition deployed by computer scientists and AI experts themselves.\footnote{We should note that there are at least two more similar shifts to contend with here: the notions of containerization and streaming, with the latter being closely linked with pipelining. Both of these have started to be studied by scholars in this domain, including in some of the current research of the authors here.} This is a world made by, and perhaps even sustained by, stacks, no doubt about it, but for this specific situation, we suggest that our methodologies might be better served by alternative conceptualizations and analytics of the shows and flows of contemporary AI systems. Furthermore, for critical AI studies, one can imagine the buffet of possibilities that a materially rooted concept such as pipeline proffers. In a fast and fluid sociotechnical world, the pace, flexibility, and directional constriction/expansion offered by pipelines might offer us some interesting critical tools for plug-and-play. Some assembly required, and all hazard warnings as per legal advice apply, of course.

\section*{What is to be done?}

What is to be done? We mean to invoke this heavy question with some sense of levity but also an equal amount of seriousness. The methodological problems identified above are by no means exhaustive, but they point to a critical lacuna in critical scholarship on AI. Some of the issues stem from a mismatch between the object of critique and the state of methodological scholarship, others from rhetorical gaps in templated, programmatic discourse, and yet others from a shifting technical terrain that temporally exceeds the bounds of movement of scholarly inquiry. Regardless of the origins, however, we firmly believe that some methodological reflection within the field at large is at order, and is possible.

Perhaps the most important reflective moment pertains simply to the alignment (and we use this word critically referencing the amorphous sense of alignment within the tech sector) between the audience and the argument. If the primary takeaway of critical scholarship is some version of `technology is bad', then it begs the question: who is this argument for? The scholars working in critical AI studies are already likely to be of the persuasion that technology is political and that contemporary politics of technological culture skews regressive and exploitative in demographically differential formations; this is, quite simply, the 101, the ground truth of our subdiscipline, and one that is unlikely to be helped by more theoretical application of older methodologies \autocites{Golumbia_2024, Chun_2008, Dhaliwal_2023}. Sure, there are other audiences worth winning, and several important lines of inquiry within this frame yet to be explored, but by and large, far too often, we—and we hope we have made it clear that this piece is self-critical first and foremost, so we as authors absolutely implicate ourselves within the collective ‘we’—are preaching to the choir.

What is to be done, then, instead? We would like to argue that multiple steps are necessary to build the foundations for a critical AI methodology -- steps we can only hint at here, leaving their implementation (or refutation) as an exercise to the reader, or rather to the critical AI studies community at large. First is an increased openness towards the technical disciplines, taking inspiration, perhaps, from earlier calls for a better integration of philosophical critique and technical development, such as in the work of Phil Agre during the previous artificial intelligence hype cycle \autocite{Agre_1997}. Technical benchmarks have progressed beyond the Turing test, as we have outlined above, and so should the methodology of critical AI studies. Second, a ‘humanities’ way of benchmarking would actually need to return to the roots of humanities critique in the close reading of complex systems with multiple dependencies within and outside of the system (nothing else are literary texts, after all). As things stand, we are performing the techno-critical equivalent of judging a book by its cover, or rather writing it, akin to Jean Paul’s protagonist in \textit{Life of the Merry Little Schoolmaster Maria Wutz in Auenthal} who, due to being too poor to buy books, simply writes them himself, taking clues only from their titles in the Leipzig Book Fair  catalog. Third, and maybe most important, is the simple willingness to have the methodological discussion, to look left and right of our disciplinary niches, pausing, just for a moment, the fear that the simple existence of numbers and methods based on them will sabotage the humanities' very foundations.

\section*{Acknowledgements}

The authors would like to thank Anna Schewelew, Rita Raley, Thao Phan, Markus Krajewski, N. Katherine Hayles, John Behrens, Katherine Buse, and Kate Marshall, for their thoughts and conversations.

\printbibliography

\end{document}